\documentclass{aa}  
\usepackage{graphicx}
\usepackage{float}
\usepackage[export]{adjustbox}
\usepackage{txfonts}
\usepackage[usenames, dvipsnames]{color}
\usepackage{xfrac}
\usepackage{fancyhdr}
\usepackage[colorlinks=true,allcolors=blue]{hyperref}%
\usepackage{ccaption}
\usepackage{soul}

\newcommand\shorttitle{Faraday Moments of STAPS}
\newcommand\authors{Raycheva et al.}

\begin{document}

   \title{Faraday moments of the Southern Twenty-centimeter All-sky Polarization Survey (STAPS)}

   \author{N. Raycheva\inst{1},
          M. Haverkorn\inst{1},
          S. Ideguchi\inst{2},
          J. M. Stil\inst{3},
          X. Sun\inst{4},
          J. L. Han\inst{5,6},
          E. Carretti\inst{7},
          X. Y. Gao\inst{5},
          A. Bracco\inst{8},
          S. E. Clark\inst{9,10},
          J. M. Dickey\inst{11},
          B. M. Gaensler\inst{12,13,14},
          A. Hill\inst{15,16},
          T. Landecker\inst{15},
          A. Ordog\inst{15,16},
          A. Seta\inst{17},
          M. Tahani\inst{18},
          M. Wolleben\inst{19}
          }

   \institute{Department of Astrophysics/IMAPP, Radboud University, PO Box 9010, 6500 GL, The Netherlands\\ %1
              \email{n.n.raycheva@gmail.com}
        \and
            National Astronomical Observatory of Japan, 2-21-1 Osawa, Mitaka, Tokyo 181-8588, Japan%2
        \and
             Department of Physics and Astronomy, The University of Calgary, 2500 University Drive NW, Calgary, Alberta, T2N1N4, Canada%3
        \and
            School of Physics and Astronomy, Yunnan University, Kunming 650500, China%4
        \and
            National Astronomical Observatories, Chinese Academy of Sciences, 20A Datun Road, Chaoyang District, Beijing 100101, China %5
        \and
            School of Astronomy, University of Chinese Academy of Sciences, Beijing 100049, China%6
        \and
            INAF Istituto di Radioastronomia, via Gobetti 101, 40129 Bologna, Italy%7
        \and
            INAF -- Osservatorio Astrofisico di Arcetri, Largo E. Fermi 5, 50125 Firenze, Italy%8
        \and
            Department of Physics, Stanford University, Stanford, CA 94305, USA%9
        \and
            Kavli Institute for Particle Astrophysics \& Cosmology, P.O. Box 2450, Stanford University, Stanford, CA 94305, USA%10
        \and
            School of Natural Sciences, Private Bag 37, University of Tasmania, Hobart, TAS, Australia%11
        \and
            Dunlap Institute for Astronomy \& Astrophysics, University of Toronto, 50 St.~George Street, Toronto, ON M5S 3H4, Canada%12
        \and
            David A.~Dunlap Department of Astronomy \& Astrophysics, University of Toronto, 50 St.~George Street, Toronto, ON M5S 3H4, Canada%13
        \and
            Division of Physical and Biological Sciences, University of California Santa Cruz, 1156 High Street, Santa Cruz, CA 95064, USA%14
        \and
            Dominion Radio Astrophysical Observatory, Herzberg Astronomy and Astrophysics Research Centre, National Research Council Canada, PO Box 248, Penticton, B.C. Canada, V2A 6J9%15
        \and
            Department of Computer Science, Math, Physics, \& Statistics, The University of British Columbia, Okanagan Campus, Kelowna, BC V1V 1V7, Canada%16
        \and
            Research School of Astronomy and Astrophysics, Australian National University, Canberra, ACT 2611, Australia%17
        \and
            Banting and KIPAC Fellowships: Kavli Institute for Particle Astrophysics \& Cosmology (KIPAC), Stanford University, Stanford, CA 94305, USA%18
        \and
            Skaha Remote Sensing Ltd., 3165 Juniper Drive, Naramata, British Columbia V0H 1N0, Canada%19
             }

   \date{Received 2023; accepted 2024}

% \abstract{}{}{}{}{} 
% 5 {} token are mandatory

  \abstract
{Faraday tomography of broadband radio polarization surveys enables us to study magnetic fields and their interaction with the interstellar medium (ISM). Such surveys include the Global Magneto-Ionic Medium Survey (GMIMS), which covers the northern and southern hemispheres at $\sim$ 300--1800 MHz.}
{In this work, we used the GMIMS High Band South (1328--1768 MHz), also named the Southern Twenty-centimeter All-sky Polarization Survey (STAPS), which observes the southern sky at a resolution of 18$\arcmin$.}
{To extract the key parameters of the magnetized ISM from STAPS, we computed the Faraday moments of the tomographic data cubes. These moments include the total polarized intensity, the mean Faraday depth weighted by the polarized intensity, the weighted dispersion of the Faraday spectrum, and its skewness. We compared the Faraday moments to those calculated over the same frequency range in the northern sky (using the Dominion Radio Astrophysical Observatory, DRAO), in a strip of $360\degr \times 30\degr$ that overlaps with STAPS coverage.}
{We find that the total polarized intensity is generally dominated by diffuse emission that decreases at longitudes of $l \leq 300\degr$. The Faraday moments reveal a variety of polarization structures. Low-intensity regions at high latitudes usually have a single Faraday depth component. Due to its insufficiently large frequency coverage, STAPS cannot detect Faraday thick structures. Comparing the Faraday depths from STAPS to total rotation measures from extragalactic sources suggests that STAPS frequencies are high enough that the intervening ISM causes depolarization to background emission at intermediate and high Galactic latitudes. Where they overlap, the STAPS and DRAO surveys exhibit broad correspondence but differ in polarized intensity by a factor of $\sim$1.8.}
{}

    \titlerunning{short title}
    \authorrunning{name(s) of author(s)}
    \maketitle
%
%-------------------------------------------------------------------

\section{Introduction}

Knowledge of Galactic magnetism is crucial to interpreting many astrophysical processes observed in the Galaxy, such as star formation (e.g., \citealt{Li2009,Wurster2018}), cosmic ray propagation (e.g., \citealt{Minter1996,Giacalone2017}), and cloud formation (e.g., \citealt{Tahani2022}). The interpretation of diffuse polarized synchrotron emission in the Galaxy is aided by broad wavelength ranges, allowing us to use the Faraday tomography technique  \citep{Burn1966,Brentjens2005,Brentjens2011}, which can disentangle the polarized synchrotron emission through the frequency-dependent Faraday rotation it experiences along the line of sight (LoS), caused by magnetic fields.

The change in orientation of polarization angle ($\Delta\chi$) at a certain wavelength ($\lambda$) through a magneto-ionic medium in an LoS between a source and an observer due to Faraday rotation is an efficient tool with which to study the properties of the Galactic magnetic fields at radio frequencies (e.g., \citealt{Burn1966,Brentjens2005}). The observed polarization angle is defined with Stokes parameters $Q$ and $U$ as (IAU convention)

 \begin{equation} \label{eq:pa}
\chi = \frac{1}{2}\, \rm{arctan}\left(\frac{\rm U}{\rm Q} \right) {\rm (rad)}. 
\end{equation}

If a signal is Faraday-rotated, its change in polarization angle is
\begin{equation}\label{eq:pa}
\Delta\chi = 0.812\lambda^2\int_d^0 n_{e}(r) B_{\parallel}(r) dr = \lambda^2 RM,
\end{equation}
where $n_{e}$ is the free thermal electron density (cm$^{-3}$), $B_{\parallel}$ is the magnetic field component parallel to the LoS ($\mu$G), and the bounds of the integration cover the entire LoS (pc) from the source at a distance, $d$ (pc), to the observer. Here, $r$ is a location along the path length and $dr$ is an incremental displacement between a source and an observer along the LoS. If the polarization angle varies linearly with $\lambda^2$, the rotation measure (RM) can be calculated as

\begin{equation}
RM \equiv \frac{d\chi(\lambda^2)}{d\lambda^2} ~~\rm(rad~m^{-2}),
\end{equation}\label{eq:RM}

where the sign of \textit{RM} provides the direction of $B_{\parallel}$; that is, a negative [positive] sign of \textit{RM} signifies a $B_{\parallel}$ pointing away from [toward] the observer.

If linearly polarized emission and Faraday rotation are mixed along an LoS, Faraday rotation is described not by a single number, \textit{RM}, but as a distance-dependent Faraday depth ($\phi$),
\begin{equation}\label{eq:rm}
\phi(d) \ = \ 0.812\intop_{d}^{0}n_{e}(r)B_{\parallel}(r)dr ~~\rm(rad~m^{-2}).
\end{equation}
The difference between this equation and Eq.~\ref{eq:pa} is that the distance, $d$, to the emitting volume is variable, resulting in a spectrum of Faraday depths rather than a single RM value.

The RM-synthesis algorithm is a technique utilized to analyze and extract information about the observed polarized intensity (PI), \((P(\phi) = \sqrt{Q(\phi)^{2} + U(\phi)^{2}})\), by reconstructing the distribution of polarized emission as a function of $\phi$ \citep{Burn1966,Brentjens2005}. The output Faraday spectrum is deconvolved to recover the underlying polarized emission profile and separate the contributions from different Faraday depths. The complex mixture of Faraday rotations can be disentangled using this process. $RM$-synthesis computes a unique sampling function representing the finite frequency coverage, called the \textit{RM} spread function (R($\phi$), RMSF), for each image pixel in a data cube, which characterizes the resolution of the observation in $\phi$. The RMSF is convolved with the true Faraday spectrum. It can be deconvolved, using algorithms like \texttt{RM-CLEAN} for partial correction of distortion from the dirty beam \citep[see][for details]{Heald2009, Kumazaki2014}. We used RM synthesis to perform Faraday tomography, which allows us to examine the complex distributions of magnetized and ionized gas in the Faraday spectrum along each LoS.

As was described by \citet{Brentjens2005} in equations 61--63, $RM$-synthesis has limitations. The full width at half maximum of the main lobe of R($\phi$), the Faraday resolution, is \citep{Dickey2019}
\begin{equation} \label{eq:fwhm}
\delta \phi = \frac{3.79}{\Delta\lambda^{2}},
\end{equation}
where $\Delta(\lambda^{2})$ is the squared wavelength range of the observing bandwidth. The maximum Faraday depth that can be detected is given by
\begin{equation} \label{eq:phi_max}
\phi_{max} \ = \ \frac{1.9}{\delta\lambda^2},
\end{equation}
where $\delta\lambda^2$ is the squared wavelength range of each channel. Additionally, the broadest feature that can be detected is given by the squared minimum wavelength ($\lambda_{min}^2$) as
\begin{equation} \label{eq:phi_max_scale}
\phi_{max-scale} \ \simeq \ \frac{\pi}{\lambda_{min}^2}.
\end{equation}

We can see two observational features within Faraday spectra: Faraday simplicity and Faraday complexity \citep{Brentjens2005}. A Faraday simple spectrum contains one component, indicating a synchrotron emitting source with Faraday rotation (by a `Faraday screen') between the source and observer \citep{Brentjens2005}. A Faraday complex spectrum occurs when polarized emission and Faraday rotation are mixed along the LoS (or within the observation beam). In terms of observations, a feature is Faraday thick if it is broad compared to the Faraday resolution, and the ability to detect a Faraday thick structure depends on the observing wavelength, which is specifically limited by Eq.~(\ref{eq:phi_max_scale}). These sources extend in the Faraday spectrum over a width, $\Delta\phi$, which occurs when \(\lambda^2\Delta\phi\gg1\). As is seen in the wavelength dependence of the thickness of a Faraday spectrum, the higher the frequency, the broader the minimum observable width of the Faraday spectrum.

Visualization and interpretation of data cubes produced by Faraday Tomography can be enhanced by moment mapping \citep{Dickey2019}. This involves the integration of various quantities (weighted by the PI) along the Faraday depth axis, which produces a scalar quantity per spatial pixel known as a statistical Faraday moment. The analysis of statistical moment maps simplifies the understanding of some main characteristics of the Faraday cubes and, by extension, interstellar magnetic fields.

An observational project in the last two decades, known as the Global Magneto-Ionic Medium Survey (GMIMS), has given us a better understanding of the role of magnetic fields in the interstellar medium (ISM). It measures diffuse polarized emission from approximately 300 MHz to 1800 MHz across the entire sky, probing both the southern \citep{Wolleben2009,Wolleben2019} and northern \citep{Wolleben2021} hemispheres. Many studies have already drawn attention to the technique of moment mapping using radio polarization surveys. \citet{Hill2018} modeled a dense ionized ISM component embedded in warm ionized diffuse gas and calculated Faraday spectra and moments at various GMIMS bands. He found that depolarization at larger distances was lower than expected, and observed that Faraday depths could even be randomized. \citet{Thomson2018} studied the $B_\parallel$ of the Galactic supershell GSH 006$-$15$+$7 using the S-band Polarization All Sky Survey (S-PASS) and generated maps of the first moment for the parameters of the Faraday rotation model and Faraday thickness. \citet{Wolleben2019} mapped the zeroth and first moments of the southern sky at 300$-$480 MHz with the 64m Parkes telescope \textit{Murriyang} as part of the GMIMS Low-Band survey. \citet{Dickey2019} presented a detailed comparison between the moments of this survey and the GMIMS-North High-band survey, taken with the 26m Dominion Radio Astrophysical Observatory (DRAO) telescope at 1270$-$1750~MHz.  \citet{Thomson2021} used the same data as \citet{Dickey2019} to study the Faraday moments of the G150$-$50 region in detail. Inspecting its magnetic field properties, they found that this region is related to Radio Loop II. For a statistical examination in multiphase ISM, \citet{Bracco2022} computed moment maps on the Faraday cubes of mock observations of colliding supershells generated by stellar feedback with frequency ranges of 115--170 MHz adapted for LOFAR observations (see references therein). They found that radio sky observations at low frequencies require multiphase modeling approaches to magnetohydrodynamic processes. Using moment maps from the LOFAR Two-meter Sky Survey, \citet{Erceg2022} found a correlation with the extragalactic RM of \citet{Hutschenreuter2022} and with the DRAO GMIMS at 1.4 GHz \citep{Dickey2019} in the Loop III region. \citet{Snidaric2023} also used the moments of the LOFAR data to inspect the ELAIS-N1 field and found a connection to the polarized emission from Loop III.

This paper studies Faraday depth computed from GMIMS High Band South (GHBS), also known as the Southern Twenty-centimetre All-Sky Survey (STAPS). The survey covers 1328$-$1768~MHz. For data verification, we compare it with moment maps from GMIMS High Band North (GHBN, \citealt{Wolleben2021}) in the sky common to both surveys. Here, we only focused on the survey data in Faraday depth space, and Sun et al. (in prep.) present the general survey details and overview.

This study is organized as follows. We present the details of the data used in Sect.~\ref{sec:data}. The techniques to calculate the moment maps from the Faraday cubes of STAPS are explained in Sect.~\ref{sec:method}. We present the Faraday moments of STAPS in Sect.~\ref{sec:staps_mom}. We give our data verification results in Sect.~\ref{sec:data_ver}. We discuss our findings in Sect.~\ref{sec:disc}. Finally, Sect.~\ref{sec:conc} presents the summary and conclusions.

\begin{table*}
\begin{center}
\caption{\label{tab:surveys}Survey properties of STAPS and GHBN.}
\begin{tabular}{lll}
\hline \\[-1.0ex]
                                     & STAPS                              & GHBN   \\
\hline \\[-1.0ex]
Declination range                    &  $\delta<0\degr$                    & $-30\degr < \delta < +87\degr$ \\[0.15 cm]
                     
Angular resolution                   & 18$\arcmin$                        & 40$\arcmin$  \\[0.15 cm]
                     
Frequency range                      & 1328 $-$ 1768 MHz                    & 1280 $-$ 1750 MHz \\[0.15 cm]

Frequency resolution                 & 1~MHz                              & 1~MHz \\[0.15 cm]
                  
$\lambda^2$ range                    & 0.029 m$^2$ $-$ 0.051 m$^2$  & 0.029 m$^2$ $-$ 0.055 m$^2$  \\ [0.15 cm]
                     
$\Delta \lambda^2$                   & 0.022 m$^2$                        & 0.026 m$^2$ \\[0.15 cm]
                     
$\delta \lambda^2$                   & (3.2 $-$ 7.7) $\times$ 10$^{-5}$ m$^2$                                  & (3.3 $-$ 8.6) $\times$ 10$^{-5}$ m$^2$ \\[0.15 cm]

RM resolution ($\delta \phi$)          & 170~rad~m$^{-2}$       & 148~rad~m$^{-2}$\\[0.15 cm]

RM range ($\phi_{max}$)               & (2.2 $-$ 5.3) $\times$ 10$^4$ rad~m$^{-2}$                                   & (2.1 $-$ 5.2) $\times$ 10$^4$ rad~m$^{-2}$ \\[0.15 cm]

RM feature width ($\phi_{max-scale}$)  & 1.1 $\times$ 10$^2$ rad~m$^{-2}$    & 1.1 $\times$ 10$^2$ rad~m$^{-2}$\\ [0.15 cm]

Cleaned $\phi$ spectral range (for the entire survey)        & $\pm$750 rad~m$^{-2}$             & $-$ \\ [0.15 cm]

Cleaned $\phi$ spectral range (for the overlap region)        & $\pm$500 rad~m$^{-2}$             & $\pm$500 rad~m$^{-2}$ \\ [0.15 cm]

$\phi$ sampling (for the entire survey)      & 5 rad~m$^{-2}$                     & $-$  \\[0.15 cm]

$\phi$ sampling (for the overlap region)      & 1 rad~m$^{-2}$                     & 1 rad~m$^{-2}$  \\[0.15 cm]

 \hline
\end{tabular}
\end{center}
\end{table*}

\section{Data}\label{sec:data}

\subsection{Southern Twenty-centimetre All-Sky Survey}
The data presented here are part of the GMIMS project \citep{Wolleben2009}. This survey was started in 2008 as an independent project named STAPS and observed mostly piggybacking on its sister survey, S-PASS \citep{Carretti2019}, which covered the frequency range of 2176$-$2400~MHz. Later, STAPS was incorporated into GMIMS as the GMIMS High Band South survey. 

The STAPS observations were made with the 64m Parkes telescope \textit{Murriyang} using an H-OH receiver. The S-PASS Galileo receiver was placed in prime focus, and the H-OH receiver was located off-axis next to the Galileo receiver, displaced 630~mm from the prime focus horizontally, and 7.6~mm vertically to keep the feed in focus. In this way, STAPS was able to piggyback on the S-PASS survey without requiring additional observing time, with some penalty in the form of beam distortions (see below). The complete survey and data products will be presented by Sun et al. (in prep).

The Southern Twenty-centimetre All-Sky Survey covers the frequency range of 1328$-$1768 MHz with a frequency resolution of 1 MHz, and the whole sky south of a declination of $0{\degr}$. The data reduction and map-making process were analogous to the data processing and mapping of the S-PASS survey \citep{Carretti2019}: the scanning was done using fast azimuth scans at a constant elevation of \(EL=33{\degr}\), the elevation of the South Celestial Pole as seen from Parkes. Scanning at this elevation allows data to be gathered as far south in declination as $-90\degr$. The northern limit was set by choosing the azimuth coverage of the scans. The advantage of scanning at a constant elevation is that ground emission is constant. The scans were as long as possible ($\sim$120${\degr}$) to reduce the chances of missing large-scale structures and to perform a polarization calibration using the parallactic angle modulation. The complete southern sky was scanned both toward the east and the west, after which the east and west scans were basket-woven into a sky map. This technique was developed for S-PASS (see \citealt{Carretti2019} for details).

The offset position of the feed led to a coma lobe, and a concomitant broadening of the main beam, together with a decrease in the forward gain of the telescope. The peak strength of the coma lobe is about 6\% to 9\% of the total intensity from the highest to the lowest frequency. As the angular location of the coma lobe varies per observing beam, its effect on the diffuse polarized emission is thought to be between 2\% and 4\% off-axis, increasing with frequency (Sun et al., in prep). 
The STAPS total-intensity scale is tied to the GMIMS HBN total-intensity scale, which is absolutely calibrated. In STAPS, the PI scale is tied to the total-intensity scale, as is detailed in the survey paper (Sun et al., in prep). Here, the polarization data were convolved to a common resolution of 18$\arcmin$, assuming a Gaussian observing beam. The maps in Jy~beam$^{-1}$ were converted into K using the conversion factor of \(0.537\times(\nu_{0}/\nu)^{2}\) K/Jy, where $\nu$ is any frequency in the coverage, and $\nu_{0}$ is 1400 MHz. This conversion ignores the partial redistribution of intensity outside the main beam (predominantly into the coma lobe), which overestimates the intensity scale in K. Observational details that are relevant to this paper and parameters of the Faraday cubes are given in Table~\ref{tab:surveys}.

\subsection{The Global Magneto-Ionic Medium Survey High Band North}
The GHBN survey \citep{Wolleben2021} in the northern sky covers approximately the same frequency range as STAPS, with a resolution of 40$\arcmin$. As GHBN extends down to declinations of $-30{\degr}$, there is an overlapping sky region of $\sim$30${\degr}$ in declination, which is observed by both the northern and the southern surveys. The Faraday cubes and Faraday moments have been analyzed by \citet{Dickey2019}, \citet{Ordog2020} and \citet{Dickey2022}. In this paper, we compare the Faraday cubes and moments of the GHBN survey with those of the GHBS. (see Table~\ref{tab:surveys} for the observational parameters of the GHBN survey).

\section{Producing Faraday spectra}\label{sec:method}
We now describe the computations of RM synthesis, noise map, and Faraday moments, including any preprocessing steps undertaken for analysis.

\subsection{Rotation measure synthesis}\label{sec:rm-synthesis}
We used the RM-Tools package\footnote{\url{https://github.com/CIRADA-Tools/RM-Tools}, v1.1.1} from the Canadian Initiative for Radio Astronomy Data Analysis (CIRADA) \citep{Purcell2020} to perform RM synthesis \citep{Brentjens2005} and \texttt{RM-CLEAN} \citep{Heald2009} to calculate the Faraday depth spectra using Stokes $Q$ and $U$ data cubes. The resulting Faraday depth cubes cover the Faraday depth range of $\pm$1000 rad~m$^{-2}$ in steps of 5~rad~m$^{-2}$.

\subsection{Estimation of the \texttt{RM-CLEAN} threshold}\label{sec:noise}
As the reliability of detecting the Faraday complexity of the spectrum depends on the choice of RM-CLEAN threshold \citep{Anderson2015}, this threshold needs to be carefully estimated. We chose an RM-CLEAN threshold based on the noise per pixel calculation, as is explained below.

The \texttt{RM-CLEAN} threshold (intensity threshold) has been chosen in various ways in earlier GMIMS surveys. \citet{Wolleben2019} used 0.06~K as their \texttt{RM-CLEAN} threshold, which is the survey rms. \citet{Ordog2020} used the Faraday spectra themselves to estimate the \texttt{RM-CLEAN} threshold. She calculated noise ($\sigma$) as the mean polarized emission at a Faraday depth of $|\phi| \in [500,1000]$~rad~m$^{-2}$, where no real signals are expected, to arrive at a $5\sigma$ threshold level of 0.03~K. As the emission varies greatly across the map, we decided to evaluate a pixel-by-pixel \texttt{RM-CLEAN} threshold instead.

Assuming that the noise in Stokes $Q$ and $U$ is Gaussian, the noise in the PI distribution as a function of Faraday depth, for each pixel, should follow a Rayleigh distribution \citep{Hales2012,VanEck2018}, which is a special case of the Rice distribution \citep{Rice1945}: 
\begin{equation}
R(x;\sigma)=\frac{x}{\sigma^2}e^{-x^2/2\sigma^2}, 
\end{equation}
where $0\leq x<\infty$, and $\sigma>0$ defines the noise limit. Therefore, for every pixel we fitted the distribution of PI over the Faraday depth with a Rayleigh distribution, using only Faraday depths where we expected no signal and only noise. We fitted Rayleigh distributions to each pixel in the range of $|\phi| \in [500,1000]$~rad~m$^{-2}$ and also in the range of $|\phi| \in [750,1000]$~rad~m$^{-2}$ and mapped the corresponding $\sigma$ values. The map calculated in the range of $|\phi| \in [500,1000]$~rad~m$^{-2}$ gave clearly higher $\sigma$ levels than the one calculated in $|\phi| \in [750,1000]$~rad~m$^{-2}$. We also performed a Kolmogorov-Smirnov goodness-of-fit test per pixel on each map. The results show a better fit between the data and the Rayleigh distribution in the range of $|\phi| \in [750,1000]$~rad~m$^{-2}$ than those in the range of $|\phi| \in [500,1000]$~rad~m$^{-2}$. These indicate that the latter map contains more signal (likely in side lobes that were not cleaned perfectly) than the former. As a result, we included the range of $|\phi| \in [750,1000]$~rad~m$^{-2}$ in our estimation of $\sigma$.

\begin{figure*}
\center
  \includegraphics[width=16cm,height=10cm]{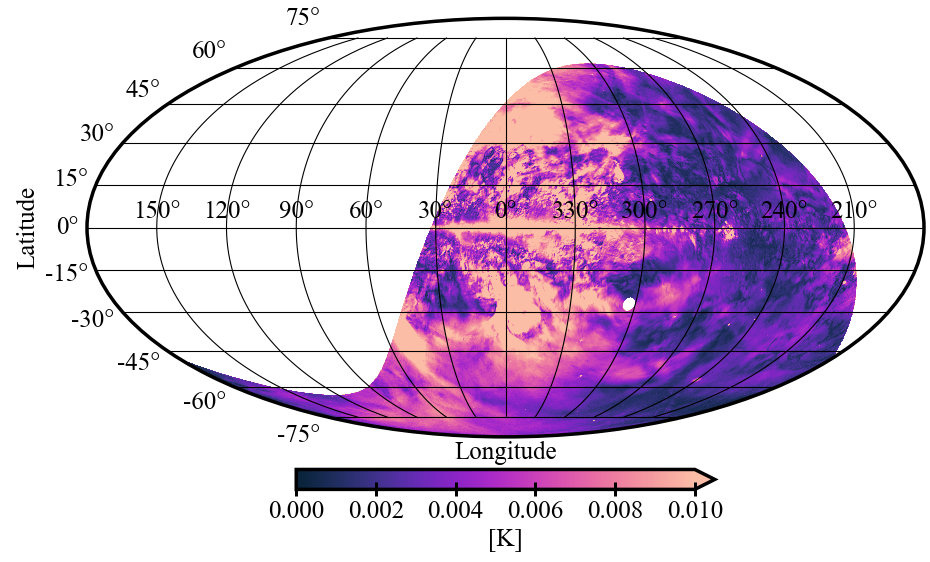}
 \caption{$\sigma$ map of STAPS showing the estimated noise level for each pixel in Galactic coordinates, shown in the Mollweide projection with the Galactic center at the center, and calculated in the range of $|\phi| \in [750,1000]$~rad~m$^{-2}$ using the technique described in Sect.~\ref{sec:noise}. The color bar is saturated to 0.01 K and the maximum is 0.2 K.}
  \label{fig:staps_noise}
\end{figure*}

The resulting map of $\sigma$ is shown in Fig.~\ref{fig:staps_noise}. It shows the 1$\sigma$ intensity threshold for each pixel. As is seen in this map, it contains features of the PI map (Sun et al., in prep), which means that in addition to instrumental noise, there are low-level side lobes of signals at lower absolute Faraday depths. We used this map to estimate the \texttt{RM-CLEAN} threshold. Since the level of side lobes in the range of $[750,1000]$~rad~m$^{-2}$ is up to 10\% of the peak intensity ($P_{peak}$, the Faraday depth in steps of 5 rad~m$^{-2}$ that has the highest $P$ in the spectrum) in Fig.~\ref{fig:staps_peakpi}, the noise is seen to be correlated with the peak intensity. We did not use a Ricean polarization bias correction to reduce the noise contributions to a polarized brightness. This is because we applied a 6$\sigma$ intensity threshold for moment map computations, at which debiasing is negligible \citep{Wardle1974,George2012}.

\begin{figure*}
\center
  \includegraphics[width=15cm]{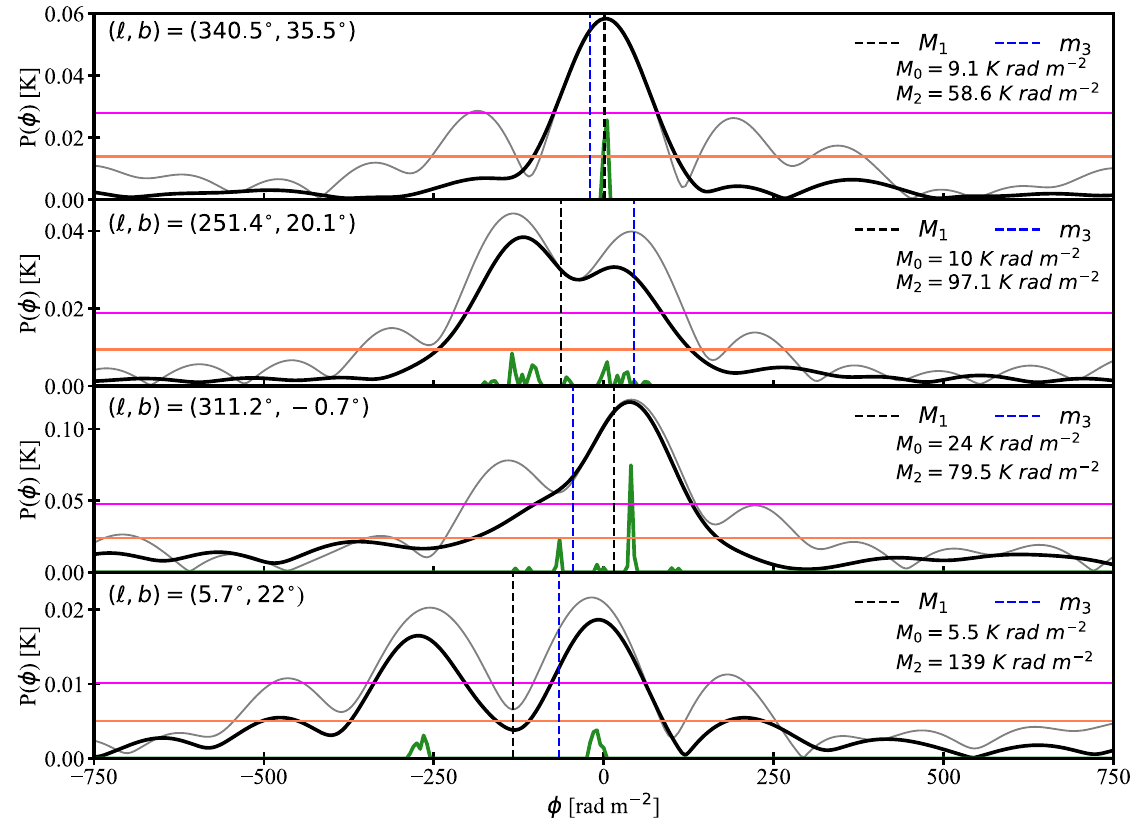}
 \caption{Faraday spectra of four sample pixels in STAPS. The orange line represents the 3$\sigma$ \texttt{RM-CLEAN} cutoff limit. The purple line shows the 6$\sigma$ intensity limit applied during Faraday moment calculations. The gray line represents the dirty Faraday depth spectrum. The black line presents the \texttt{RM-CLEAN}ed Faraday depth spectrum, and the \texttt{RM-CLEAN} model components (CC) are shown in green. The Galactic coordinates of the pixels from which the spectra were taken are shown in the top left of the panels. The legends in the top right show the Faraday moment results for the related pixels. We note the different scales on the $y$ axis in each plot.} 
  \label{fig:staps4reg}
\end{figure*}

\subsection{\texttt{RM-CLEAN}}\label{sec:rmclean}
In the \texttt{RM-CLEAN} process, we adopted a loop gain of 0.1, 1000 iterations, and a 3$\sigma$ \texttt{RM-CLEAN} threshold per pixel, using the $\sigma$ map. Fig.~\ref{fig:staps4reg} displays the dirty and clean Faraday spectra of four pixels chosen from representative regions, selected to show different spectra characteristics in the survey. The first spectrum in the upper panel consists of a single peak with one \texttt{RM-CLEAN} model component. The second panel's complex spectrum contains two low-intensity, unresolved peaks. The third panel has a resolved spectrum, broader than the RMSF, with a tail toward negative $\phi$ and with many \texttt{RM-CLEAN} model components. The fourth panel displays a double-peaked spectrum in the direction of the H\,{\sc ii} region Sh~2$-$27 \citep{Sharpless1959} at $(l, b)\sim(5\fdg7, 22\degr)$. The figure shows that side lobes are significantly suppressed by the \texttt{RM-CLEAN} process. The maximum width of Faraday depth features, $\phi_{max-scale}$, in STAPS is smaller than the Faraday depth resolution, $\delta \phi$, as is the case with GHBN, as \citet{Wolleben2010} noted. Therefore, Faraday thick structures will not be detectable. A Faraday thick structure with sharp edges may show up as two seemingly unrelated features at the edges, such as the fourth panel of Fig.~\ref{fig:staps4reg}. However, these features could also be two unrelated and unresolved emission features. Depending on the shape of the feature's edge, a Faraday-thick structure can also look like a slightly resolved component, as in the third panel.
 
\subsection{Faraday moments}\label{sec:Farmom}
We computed four moments of the \texttt{RM-CLEAN}ed image cubes by collapsing the Faraday depth cube into a Faraday moment image (\citealt{Dickey2019}) using a cutoff, which is discussed in Sect.~\ref{sec:mom_cutoff}: the total polarized brightness integrated over the Faraday spectrum (the zeroth moment, $M_0$); the mean Faraday depth weighted by the PI (the first moment, $M_{1}$); the standard deviation or width of the Faraday spectra weighted by the PI (the second moment, $M_{2}$); and the skewness (shape asymmetry) of Faraday spectra weighted by the PI (the third moment, $M_3$). For the discrete distribution of $P(\phi)$, these maps are respectively defined as \citep{Dickey2019}

\begin{equation}
        M_0 \ = \ \Delta\phi \sum_{i=1}^{n}P_i ~~~ \rm (K~rad~m^{-2}~RMSF^{-1}),
        \label{eq:mom0}
\end{equation}

\begin{equation}
        M_1 \ = \ \frac{\Delta\phi \sum_{i=1}^{n}P_i \phi_i}{M_0} ~~~ \rm (rad~m^{-2}),
        \label{eq:mom1}
\end{equation}

\begin{equation}
        M_2 \ = \ \frac{\Delta\phi \sum_{i=1}^{n}P_i \left( \phi_i - M_{1} \right)^2 }{M_0} ~~~ \rm (rad~m^{-2})^2,
        \label{eq:mom2}
\end{equation}
and
\begin{equation}
        M_3 \ = \ \frac{\Delta\phi \sum_{i=1}^{n}P_i \left( \phi_i - M_{1} \right)^3 }{M_0} ~~~ \rm (rad~m^{-2})^3,
        \label{eq:mom3}
\end{equation}

where $n$ is the number of $\phi$ values (the length of the $z$ axis in the image cube), $\phi_i$ is $\phi$ of the $i$th Faraday depth component, and $\Delta\phi$ is the step size in the $\phi$ interval \((|\phi_{i}-\phi_{i+1}|)\). $M_2$ shows the spread of the PI from $M_1$, and, following \citet{Dickey2019}, we also use its square root-normalized version (\(m_2 = \sqrt{M_2}\)). Here, $M_3$ demonstrates a Faraday spectrum's asymmetry about its mean, where $M_3 = 0$ for a symmetrical $|F|(\phi_i)$ distribution. Again, as in $m_2$, we used its root-normalized version (\(m_3 = \sqrt[3]{M_3}\)). $m_3 > 0$ indicates positive [right] skewness, while $m_3 < 0$ indicates negative [left] skewness. Throughout this paper, all plots and explanations involving second and third moments use $m_2$ and $m_3$, respectively, which have units of rad~m$^{-2}$.

For a spectrum with a single Faraday-thin peak, $M_1$ is $\phi$, at which this peak polarization happens ($M_1 = \phi_{peak}$, Fig.~\ref{fig:staps_peakrm}), $m_2$ is the width of the peak, and $m_3$ is zero. At STAPS frequencies, this is a possible scenario for high Galactic latitudes. However, $M_1 \neq \phi_{peak}$ and typically $m_2$ have high values for more complex spectra with multiple emission components.

\subsection{Faraday depth cutoff in moment maps}\label{sec:mom_cutoff}
A cutoff in the Faraday depth range needs to be applied, to exclude Faraday depths dominated by noise as much as possible in the Faraday moment calculations. This cutoff in the Faraday depth range is commonly defined as the Faraday depth range within which a pixel has an intensity above a certain level, with additional criteria to consider Faraday complexity.

As \citet{Dickey2019} mentioned, at high $\pm\phi$, noise or spurious features in the $P$ distribution strongly affect $M_1$ and $m_2$, which becomes more drastic in higher moments. Thus, they restricted the range of $\phi$ channels according to some Faraday moment map calculation criteria.  They used the following Faraday depth threshold methods for GHBN and GMIMS Low Band South (GLBS) Faraday spectra. It was calculated that each pixel should be within the intensity threshold of 5$\sigma$ (0.04~K), or 15\% of the Faraday spectrum peak, whichever is the largest, and only Faraday depth channels meeting these criteria were included. However, as they found second peaks in GLBS, they added another criterion to thresholding by fitting a Gaussian to the highest peak in the concerned spectrum and subtracting it from the entire data of that spectrum to find the second peak and again only included the Faraday depths that meet this criterion. Studying GHBN, \citet{Ordog2020} and \citet{Wolleben2021} based their criteria for thresholding on \citet{Dickey2019}, although they did not apply any Gaussian fitting to locate multiple peaks. Additionally, they ensured that the artificial features found in Faraday spectra as peaks above their 5$\sigma$ \texttt{RM-CLEAN} threshold (0.03~K) would be eliminated by adding a 33\% margin to the \texttt{RM-CLEAN} threshold (0.04~K). 

In some Faraday spectra of STAPS, there are multiple peaks (more than two) well above the 3$\sigma$ level per pixel after the \texttt{RM-CLEAN} process. Consequently, in our situation, the methods described above could cause an underestimation of the number of Faraday depth peaks, leading to the missing complexity of spectra. Hence, we did not directly utilize the exact thresholding done by \citet{Dickey2019} and \citet{Ordog2020}, but adapted them to our analysis: we used a 6$\sigma$ threshold to each pixel's intensity to minimize any possible noise peaks that can occur in the case of a polarized peak that is surrounded by high noise levels. Eventually, to compute the Faraday moments, we only incorporated the Faraday depths in the intensity peak(s) above the 6$\sigma$ threshold and masked the Faraday spectra below this threshold. During the moment map computations, we included Faraday depth channels of $\pm$500~rad~m$^{-2}$ for the overlap region, as GHBN is only reliable in this Faraday depth span (as is explained in Sect.~\ref{sec:ext_overlap}), and $\pm$750~rad~m$^{-2}$ for STAPS, high-resolution data.

\section{Faraday moment maps of the Southern Twenty-centimetre All-Sky Survey}\label{sec:staps_mom}
The Faraday moment maps of STAPS are presented in Fig.~\ref{fig:staps_moments}. In the following paragraphs, we describe each moment map.

\begin{figure*}
\center
\includegraphics[width=16cm,height=10cm]{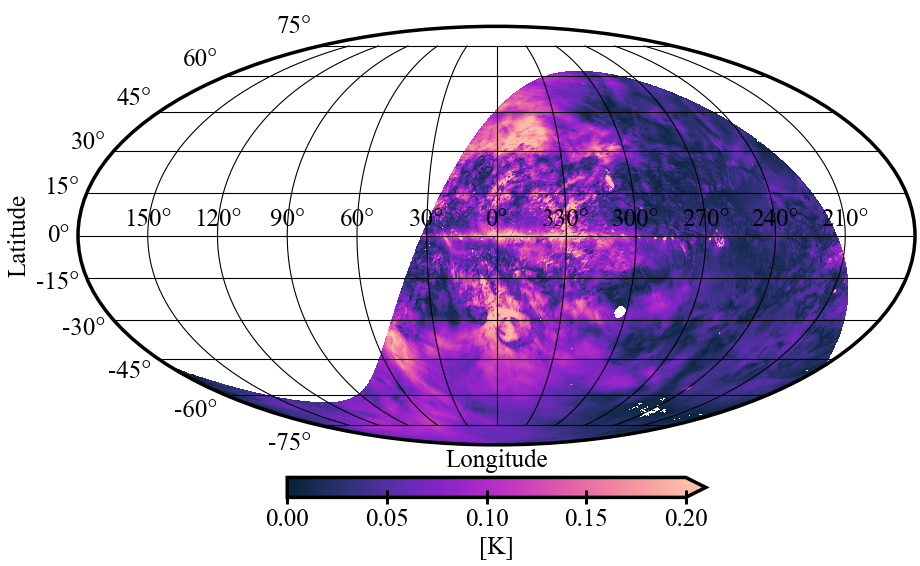}
\caption{$P_{peak}$ map of STAPS shown in Mollweide projection. The white areas are either outside the survey coverage, or masked pixels where S/N$<$6. The color bar is saturated for presentation clarity.}
\label{fig:staps_peakpi}
\end{figure*}

\begin{figure*}
\center
\includegraphics[width=16cm,height=10cm]{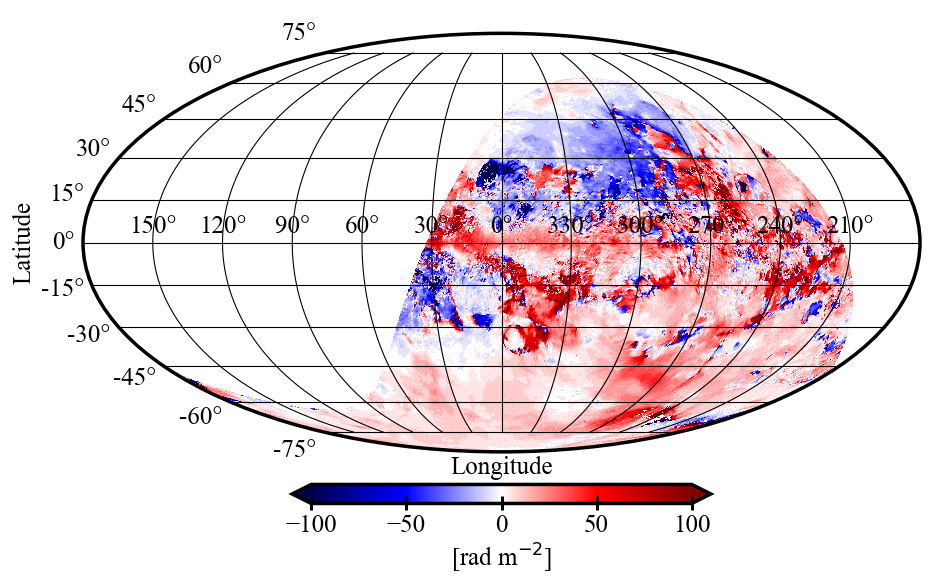}
\caption{Map of the STAPS peak Faraday depth, $\phi_{peak}$,  as fitted from the Faraday spectrum. The figure parameters are the same as in Fig.~\ref{fig:staps_peakpi}.}
\label{fig:staps_peakrm}
\end{figure*}

\begin{figure*}
\center
\includegraphics[width=16cm,height=20cm]{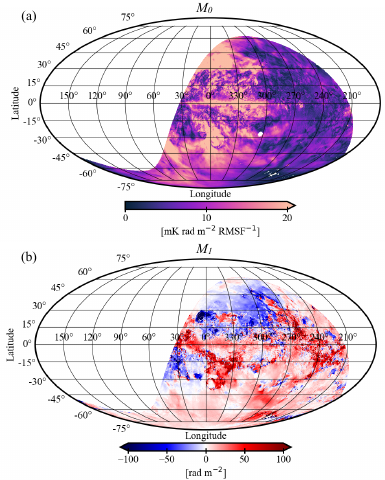}
\caption{Faraday moment maps of STAPS shown in the Mollweide projection. (a) $M_{0}$; (b) $M_{1}$; (c) $m_{2~}$(\(=\sqrt{M_2}\)); (d) $m_{3}$~(\(=\sqrt[3]{M_3}\)). The color bars are saturated to reveal the finer details. All the figure parameters are the same as in Fig.~\ref{fig:staps_peakpi}. The figure continues on the next page.}
\label{fig:staps_moments}
\end{figure*}

\begin{figure*}
\center
\includegraphics[width=16cm,height=20cm]{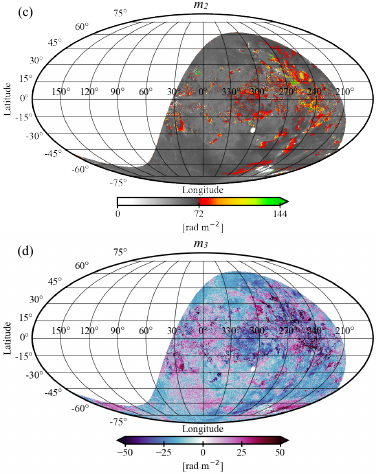}
\contcaption{Continued.}
\label{fig:continued}
\end{figure*}

In the $M_{0}$ map (Fig.~\ref{fig:staps_moments}a), there is a sharp decrease in high and/or low total polarized brightness regions: from high $M_{0}$ values (left-hand side) to low $M_{0}$ values at $l<315\degr$ (right-hand side). The high PI in the inner Galactic disk is due to instrumental polarization. Even though instrumental polarization is low (see Sect.~\ref{sec:data}), it will be significant at the location of the very high-intensity synchrotron sources in the inner Galactic disk. Moreover, we can see the Galactic center spur \citep{Sofue1989}, and the northern and southern ridges identified in S-PASS \citep{Carretti2013}. Other than these, the high $M_0$ (30~K~rad~m$^{-2}$) around the supernova remnant candidate G353$-$34 \citep{Testori2008} at $(l, b)\sim(353\degr, -34\degr)$ is quite prominent. 

The $M_{1}$ map is shown in Fig.~\ref{fig:staps_moments}(b). In the case of a unimodal unresolved Faraday spectrum, $M_{1}$ equals $\phi_{peak}$. However, the Faraday rotation sky observed with STAPS is often more complex than a single peak, except at high Galactic latitudes. In this map, the sky separates into positive and negative patches of RM over large areas, but individual structures also stand out with their $M_{1}$ values (e.g., Sh~2$-$27, G353$-$34). 

The $m_{2}$ map is shown in Fig.~\ref{fig:staps_moments}(c). It shows the width of the Faraday spectrum peaks. High values indicate a wide distribution of polarized brightness and the possibility that multiple peaks may be present in Faraday spectra; that is, Faraday complexity \citep{Alger2021}. $m_{2}$ is the smallest in an unresolved spectrum. The FWHM of the RMSF of 170~rad~m$^{-2}$ corresponds to $m_2$ for a Gaussian unresolved spectrum of $\sigma = \mbox{FWHM}/2\sqrt{2\ln 2} \approx 72$~rad~m$^{-2}$. We note that for some LoSs (shown on a gray color scale) the corresponding spectra are slightly narrower than the RMSF due to inaccuracies in the fitting.

Many diffuse structures stand out due to their high $m_{2}$ values compared to their surroundings in the $m_{2}$ map. This indicates spectral complexity, likely multiple Faraday depth components (see Sect.~\ref{sec:complexity}), such as toward the supernova remnant candidate G353$-$34.

The $m_{3}$ map in Fig.~\ref{fig:staps_moments}(d) shows the skewness of the spectrum. Two spectra that have the same $M_{1}$ and $m_{2}$ can still have different $m_{3}$, reflecting variations in their tail behaviour and overall shape. We also see certain regions separated by their skewness, such as Sh~2$-$27 with positive [right] skewness, while the higher-latitude sky contains large patches of either positive or negative skewness. The interpretation of these maps is presented in Sect.~\ref{sec:disc}.

\section{Data verification}\label{sec:data_ver}
Since the GHBN survey has a similar frequency coverage and partially overlapping declination range, we can compare the two surveys in the overlap region. Sun et al. (in prep) compare the surveys for the polarized emission in frequency space, and we compare the surveys in Faraday space in this section.

Since the two surveys cover different hemispheres, the overlap region is necessarily at the edge of both surveys. In particular, for the GHBN survey, this means a deterioration of data quality that we will have to take into account \citep{Wolleben2021}.

\subsection{Extracting the overlap region between STAPS and GHBN}\label{sec:ext_overlap}
STAPS and GHBN overlap in the region \(\delta = [-30{\degr},0{\degr}]\), which we refer to as “the overlap region” hereafter.

\begin{figure}
\center
  \includegraphics[height=9.cm]{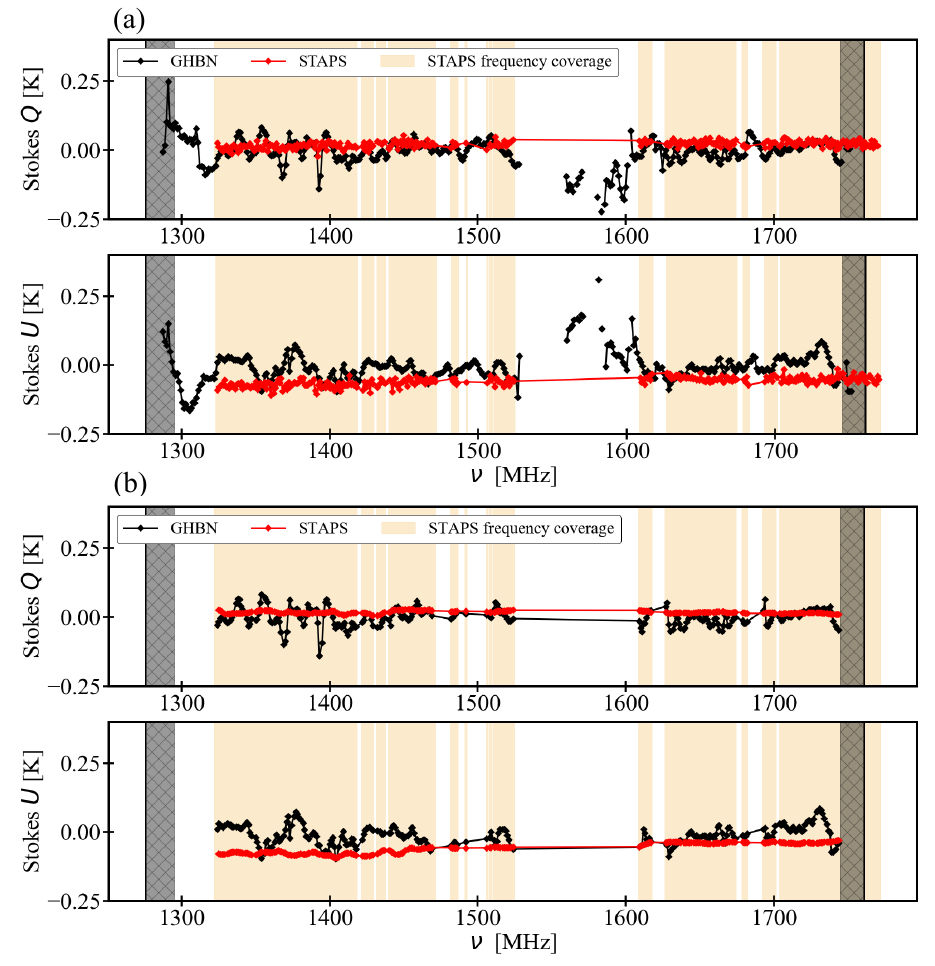}
  \caption{Representation of the frequency coverage of STAPS and GHBN at the Galactic location of \(l=310\fdg2, b=52\fdg5\). The top panel plots (a) present the original Stokes $Q$ and $U$ of STAPS and GHBN, and the bottom panel plots (b) present the data modified after the procedure explained in Sect.~\ref{sec:data_ver}. The area highlighted in yellow represents the STAPS frequency coverage. The hatched gray areas indicate the frequency ranges for which GHBN lacks adequate data in the overlap region for proper comparison. These pixels were discarded in the original data due to the localized RFI, as is described by \citet{Ordog2020}.}
  \label{fig:freq}
\end{figure}

The frequency coverage, the pixel size, the angular resolution, and the units must be the same in the overlap region to compare the two surveys. Accordingly, we applied the following steps:
\begin{enumerate}
    \item[i)] The Stokes $Q$ and $U$ maps of STAPS were smoothed to the GHBN resolution, ensuring flux conservation, and re-gridded onto the GHBN pixel grid. 
    \item[ii)] The overlap region was extracted from both surveys and projected onto equatorial coordinates.
    \item[iii)] STAPS and GHBN cover similar frequencies with slight differences (see Table~\ref{tab:surveys}), but use the same frequency channel width. Thus, we only used the frequency range common to the two surveys.
    \item[iv)] We repeated the RM-synthesis procedure and the cleaning described in Sect.~\ref{sec:rmclean} to produce new cleaned Faraday spectra and moment maps of the overlap region using the smoothed $Q$ and $U$ STAPS maps.
    \item[v)] Faraday depths in GHBN beyond $\pm$500~rad~m$^{-2}$ are dominated by spurious features, especially in the southern limit of the survey (in the area of the overlap region in Fig.~6 of \citealt{Wolleben2021}). Thus, we did not include these Faraday depth ranges in our analysis of the overlap region.
\end{enumerate}
Fig.~\ref{fig:freq} shows the results of steps 1--3 of the procedure for an example pixel that is typical of high Galactic latitudes, where both the frequency coverages are equalized and STAPS data are smoothed to the GHBN resolution. Fluctuations in STAPS data are significantly lower than in GHBN due to the higher noise in GHBN than in STAPS.

We performed RM synthesis and \texttt{RM-CLEAN} to compute the Faraday depth cubes of the overlap region. We adopted pixel-wise 6$\sigma$ cutoff parameters from the $\sigma$ map calculated for each survey to determine the Faraday depth cutoffs needed to calculate the moment maps, as is explained in Section~\ref{sec:Farmom}. The STAPS and GHBN maps are displayed in Fig.\ref{fig:overlap_all}. Unlike GHBN, the Stokes $I$ leakage into the Stokes $Q$ and $U$ are not subtracted in the current STAPS data. Since this effect is thought to be most significant in the Galactic plane, we masked the Galactic plane in the comparison. The area in GHBN most affected by ground emission, at declinations below $-25\degr$ \citep{Wolleben2021}, is also masked.

\subsection{Comparison between STAPS and GHBN}\label{sec:compDRAO}
\begin{figure*}
\center
  \includegraphics[width=16.5cm]{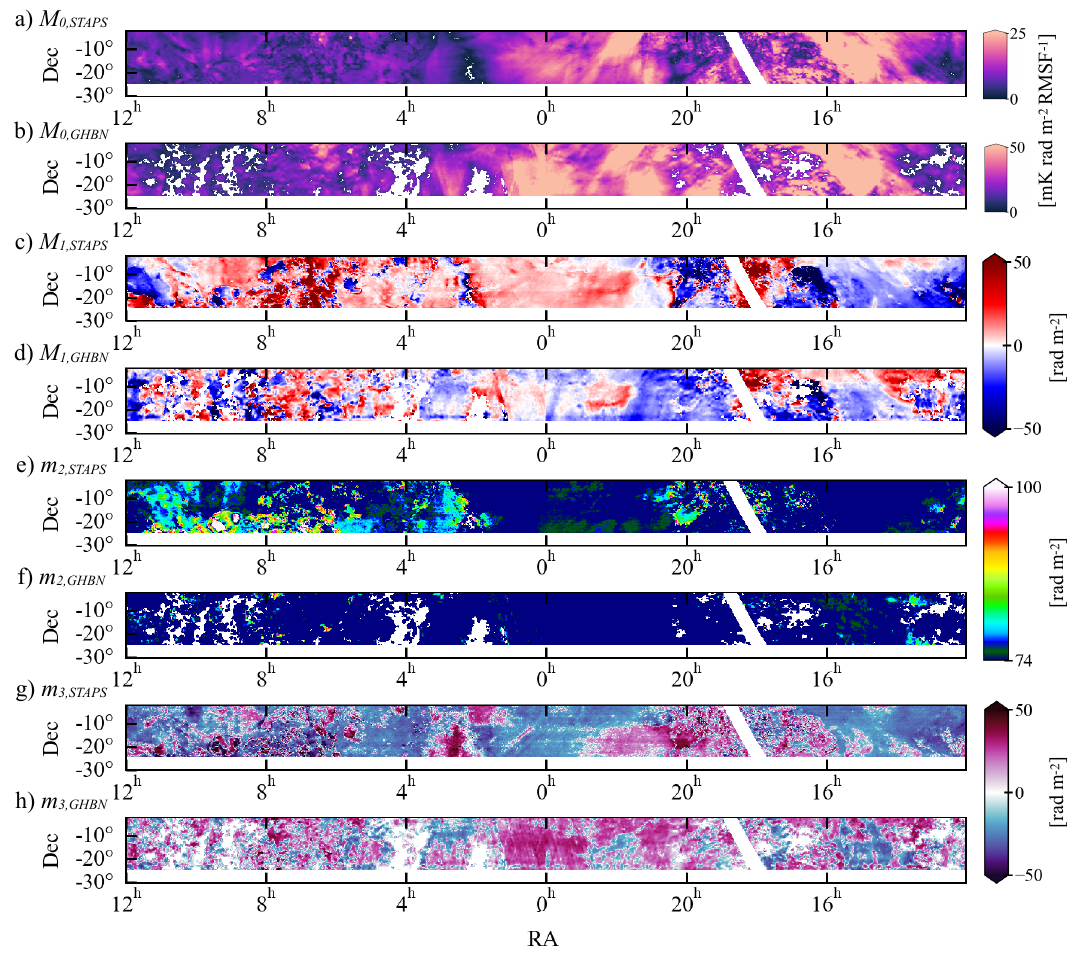}
  \caption{Moment maps of the overlap region. From top to bottom: (a) $M_{0,STAPS}$; (b) $M_{0,GHBN}$; (c) $M_{1,STAPS}$; (d) $M_{1,GHBN}$; (e) $m_{2,STAPS}$; (f) $m_{2,GHBN}$; (g) $m_{3,STAPS}$; (h)  $m_{3,GHBN}$. The white areas show either the masked values or the value of zero. The color bars are saturated to present the finer details of structures, and the $M_{0}$ color maps specifically are saturated at 50$\%$ of the maximum.}
  
  \label{fig:overlap_all}
\end{figure*}

The Faraday moment maps of the overlap region of both surveys are presented in Fig.~\ref{fig:overlap_all}. The moment maps in Figs.~\ref{fig:overlap_all}(a) and \ref{fig:overlap_all}(b) show that STAPS and GHBN show similarities in many places, such as, in the $M_0$ maps, the bright regions centered at \((\alpha, \delta) \sim (15^h, -15\degr), (22^h, -6\degr), (6^h, -4\degr)\). However, there are significant differences between the $M_{0,STAPS}$ and $M_{0,GHBN}$ maps, whereas both surveys agree on many features that are considered real Galactic signals. However, spurious signals and low signal-to-noise ratios (S/Ns) in either or both maps can easily cause significant deviations in the Faraday moment maps. Below, we discuss the agreements between the maps and the differences, with their likely causes.

\begin{figure*}
\center
  \includegraphics[width=16cm]{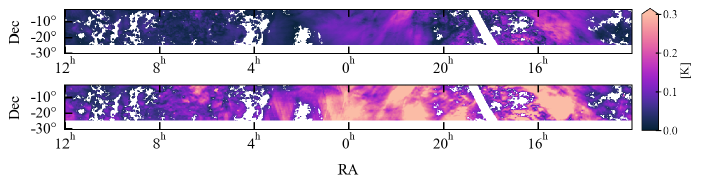}
  \caption{Maps of $P_{peak}$ of the STAPS (top) and GHBN (bottom) overlapping region. Each has the same masked values for comparison. All the figure parameters are the same as in Fig.~\ref{fig:overlap_all}.}
  \label{fig:overlap_maxPI}
\end{figure*}

\begin{figure}
\center
  \includegraphics[width=7cm,height=7cm]{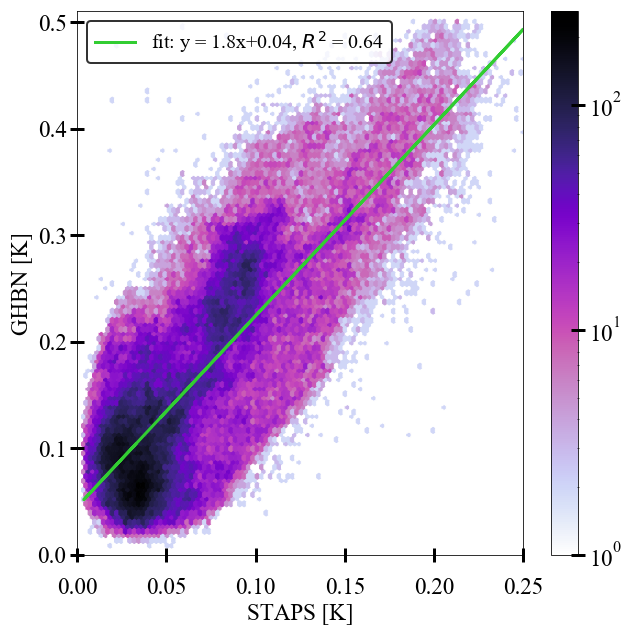}
  \caption{Scatter plot of $P_{peak}$ of STAPS and GHBN over the region where the surveys overlap. The green line corresponds to the linear fit of the data. The color bar corresponds to the number of data points in each hexbin. $R$ is the coefficient of correlation.}
  \label{fig:overlap_scatter}
\end{figure}

\begin{figure}
\center
  \includegraphics[width=8cm,height=5.8cm]{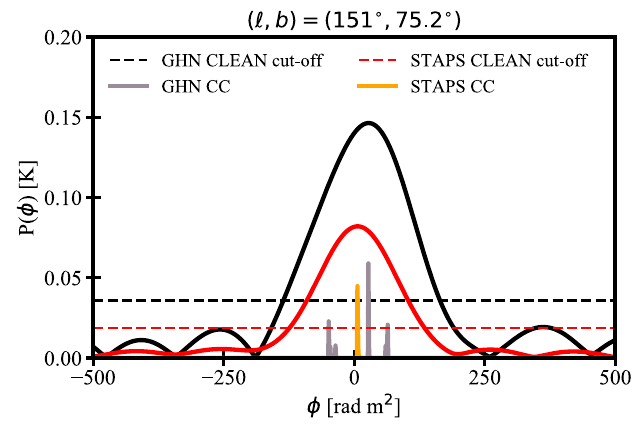}
  \caption{Representative example spectrum of the overlap region of each survey. The solid black line shows the CLEAN Faraday spectrum of the GHBN overlap region, and the solid red line shows that of STAPS.}
  \label{fig:overlap_fdf}
\end{figure}

\begin{itemize}
\item[i)] Artifacts in the GHBN maps exist that are not present in STAPS, as \citet{Wolleben2021} have already pointed out. Most notable are the zigzag patterns of the basket-weaving scanning, predominantly visible at right ascensions between about 3$^h$ and 20$^h$.
\item[ii)]The level of PI in the GHBN survey is almost twice that of STAPS. This is seen clearly in Figs.~\ref{fig:overlap_scatter} and \ref{fig:overlap_fdf}, which show, on average, a factor of 1.8 difference between the maximum PI in the Faraday spectra of the two surveys. There is evidence that the intensity scale of GHBN is generally correct. \citet{Wolleben2021} show that the total-intensity scale over the entire GHBN survey area is within 3\% of the highly reliable scale of the Stockert survey \citep{Reich1982, ReichReich1986} when main-beam brightness temperatures are compared. Additionally, \citet{Wolleben2021} compared polarized intensities over the brightest features, the Fan Region and the North Polar Spur, with polarized intensities from the Dwingeloo 1411 MHz data \citep{BrouwSpoelstra1976} and found an agreement within 10\% over both regions. The problem with GHBN polarized intensities appears to be confined to declinations below 0 degrees, getting progressively worse toward the southern limit of the GHBN survey, as is indicated by comparisons of polarized intensities between GHBN and the data of \citet{Wolleben2006} at 1410 MHz.  On the other hand, the STAPS data were compared to polarization data from the Villa Elisa 1420~MHz survey \citep{Testori2008} and found to be consistent for most of the sky. However, at high declinations, specifically at the edge of the survey's declination range, the STAPS Stokes $Q$ and $U$ signal decreased by a factor of $\sim$1.6 compared to the Villa Elisa 1420~MHz survey, indicating that the STAPS intensity may be underestimated at the edges of the survey. These two effects, together with increased artifacts at the GHBN's spatial edges and higher side lobes in the GHBN, may explain the differences in intensity between the two surveys in the overlap region. Despite this disagreement between the two surveys, we can continue to compare the higher moments, which are independent of amplitude, depending only on the change of polarization angle as a function of frequency.
\end{itemize}

The GHBN $M_1$ map is visibly affected by zigzag striping due to the scanning technique in low-intensity regions, particularly between right ascensions 3$^h$ and 20$^h$, where GHBN has more negative $M_1$ values than STAPS in that direction. A similar situation occurs between 4$^h$ and 12$^h$, which is a low-intensity region. Furthermore, around the right ascensions of 7$^h$ to 8$^h$, where the RA strip crosses the outer Galactic plane, differences between the surveys can be attributed to instrumental polarization.

The $m_{2}$ maps in Figs.~\ref{fig:overlap_all}(e) and \ref{fig:overlap_all}(f), and the $m_3$ maps in Figs.~\ref{fig:overlap_all}(g) and \ref{fig:overlap_all}(h), are also affected by the issues explained above. In general, STAPS $m_2$ is higher than GHBN $m_2$. This might be due to some real peaks in the STAPS data that may be missed in the GHBN, as it has a lower S/N than STAPS. The above-mentioned issues can cause a reversal in the skewness, as is seen in local anti-correlations of $m_3$ in STAPS and GBHN.

Fig.~\ref{fig:overlap_maxPI} shows the maps of peak PI for STAPS (top) and GHBN (bottom). As can be seen in the $M_0$ maps, an intensity difference between the two of a factor of $\sim$1.8 is also apparent in the scatter plot (Fig.~\ref{fig:overlap_scatter}). This situation is also demonstrated in the representative Faraday spectrum in Fig.~\ref{fig:overlap_fdf}, where the GHBN intensity is higher than STAPS. This intensity difference increases from a factor of 1.7 at $0^{\circ} < \delta < -10^{\circ}$ to a factor of two at $-20^{\circ} < \delta < -25^{\circ}$, supporting our conclusion that this is an edge effect.

In summary, this comparison of $PI$ and Faraday moment maps in the STAPS and GHBN surveys shows broad correspondences between the two surveys. However, in many locations, these are overshadowed by artifacts, mostly in GHBN, possibly due to incomplete removal of ground radiation (as is explained in \citealt{Wolleben2021}). Also, there is a difference in PI of about a factor of 1.8 between the two surveys. The cause of this difference is unknown, but likely related to inaccuracies because the overlap region is at the edge of the sky coverage of either survey.

\section{Discussion}\label{sec:disc}
The maps in Figs.~\ref{fig:staps_peakpi} and \ref{fig:staps_peakrm} and a comparison of the moment maps in Fig.~\ref{fig:staps_moments} illustrate properties of the magneto-ionic medium. Care has to be taken, as low-S/N regions may result in a signal in the Faraday moment maps that does not reflect structures in the ISM. For example, in the southern sky at $l \lesssim 300{\degr}$, $M_0$ mainly shows a very low PI, except for a few regions, including the Vela supernova remnant.

Comparing the first and second moment maps of Faraday depth cubes in the GMIMS Low Band South, \citep{Wolleben2019} shows no correspondence with the STAPS moment maps. As the GMIMS Low Band South has frequency coverage of 300 to 480~MHz, emission from distances greater than a few hundred parsecs is heavily depolarized, and is sensitive only to much lower Faraday depths of $\pm$10 rad~m$^{-2}$. It therefore probes only the very nearby gas \citep{Dickey2019}. Due to the lower STAPS Faraday depth resolution, STAPS is barely sensitive to the Faraday depths detected in GMIMS Low Band South. Hence, STAPS could not distinguish the multiple closely spaced peaks in some regions in GMIMS Low Band South. For the specific case of the H\,{\sc ii} region Sh~2$-$27, the $M_1$ map indicates a median RM of around $-$155~rad~m$^{-2}$, which is consistent with the value of RM found from S-PASS polarimetry in \citet{Raycheva2022} and \citet{Wolleben2021} with GHBN data.

\subsection{Complexity in Faraday spectra}\label{sec:complexity}
The $m_2$ map is an indicator of complexity in the Faraday spectra. Complexity could be caused by Faraday thick components (broad peaks), multiple Faraday thin components, or a combination. Multiple components could arise from structures along the LoS, but also within the telescope beam. Due to the frequency coverage of STAPS, the maximum detectable Faraday thickness is comparable to the resolution of Faraday depth (see Table~\ref{tab:surveys}), and the complexity of spectra will consist of multiple, potentially overlapping, unresolved peaks. In addition, distinct objects such as Sh~2$-$27 or G353$-$34 stand out, and have multiple Faraday depth peaks representing the object and the background.

\begin{figure*}
\center
\includegraphics[height=10cm]{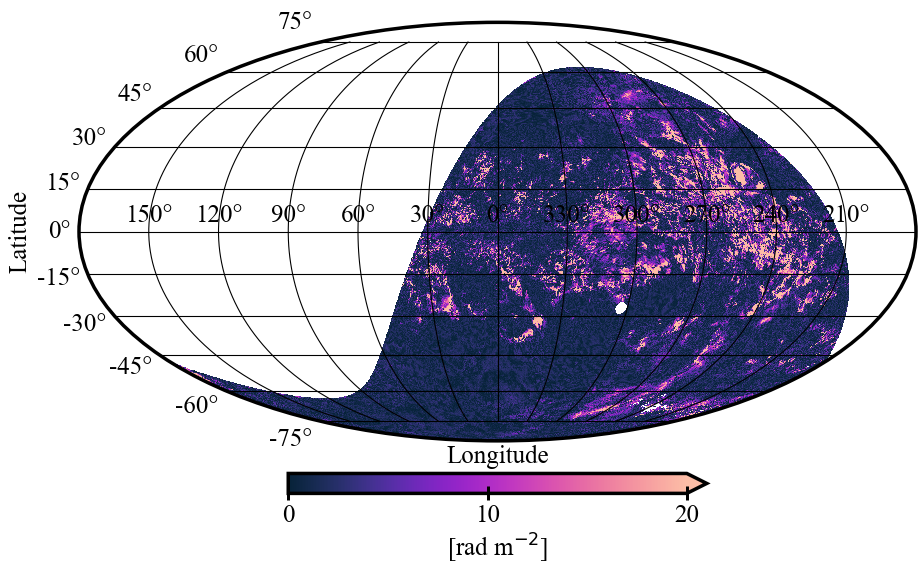}
\caption{Results of \((\phi_{peak}-M_{1})\). The parameters are the same as in Fig.~\ref{fig:staps_moments}.}
\label{fig:peakRM-M1}
\end{figure*}

\begin{figure}
\center
\includegraphics[width=8cm]{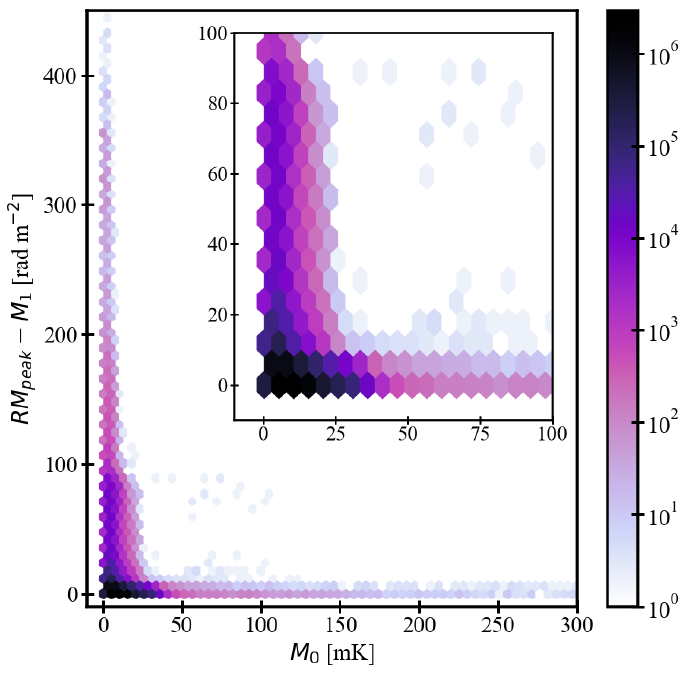}
\caption{Comparison between \(| \phi_{peak} - M_{1} |\) and $M_0$. The inner panel shows the values zoomed in between \((\phi_{peak}-M_{1})<100\) rad~m$^{-2}$ and \(M_0<100\) mK. The Galactic plane \((-6\degr<b<5\degr)\) and discrete objects such as Sh 2$-$27 and G353$-$34 are excluded. The color bar corresponds to the number of data points in each hexbin.}
\label{fig:peakRM-M1_scatter}
\end{figure}

Fig.~\ref{fig:peakRM-M1} shows the absolute difference between $\phi_{peak}$ and $M_1$ (the position of the mean relative to the peak), which is close to zero for a spectrum with a single unresolved peak or a symmetric complex spectrum, but nonzero for asymmetric complex spectra. In this figure, some parts have high $\phi_{{peak}}-M_1$ values and most of these are observed at low intensities (see Fig.~\ref{fig:staps_moments}a). As is seen in Fig.~\ref{fig:peakRM-M1_scatter}, these pixels predominantly represent regions with a low S/N. This observation indicates that in the regions where the S/N is high, the $\phi_{{peak}}-M_1$ value tends to be low, indicating the presence of single peaks, except for certain discrete areas (e.g., the Galactic plane, Sh 2$-$27, and G353$-$34). 

\citet{Bell2011} introduced and explained “Faraday caustics” as an observational signature in Faraday spectra of LoS magnetic field reversals. These caustics would be visible as Faraday depth peaks with an asymmetric tail. Although we see spectra with asymmetric tails, this asymmetry might also be due to a second overlapping Faraday depth structure. According to \citet{Bell2011}, the highest to lowest frequency ratio must be at least 1.5 to resolve this structure \citep{vanEck2019}. Since STAPS's ratio is 1.33, we are unlikely to resolve Faraday caustics. Even if we resolved Faraday caustics, these are at this resolution impossible to distinguish from blended double peaks without additional information such as high-resolution multifrequency data.

The $m_3$ map in Fig.\ref{fig:staps_moments}d can give unique information about the spectrum's shape; namely, in which direction it is skewed. In low-S/N regions, the $m_3$ map generally shows high skewness, which represents only noise. As $m_2$ and $m_3$ include increasing powers of intensity differences, the noise will dominate the low-S/N pixels in $m_2$ and especially $m_3$. Hence, according to our estimation, it is suggested that the $m_3$ map be treated with caution for signals with an S/N lower than 10. It is noticeable that most of the sky away from the plane ($|b|>30\degr$) shows a negative skewness (the tail to the left, represented in the shades of blue color), which implies an additional low-level component toward negative Faraday depths over most of the high-(negative-)latitude southern sky. Clearly, it shows the additional Faraday depth component in discrete objects. 

\subsection{Galactic center spur and S-PASS northern and southern ridges}
The large loops and spurs identified with S-PASS in \citet{Carretti2013} (such as the northern and southern ridges (extending from the center with high PI) are discernible in the STAPS $M_0$ map as well. The northern and southern ridges coincide with the high-intensity regions in $M_0$ centered on $(l, b)\sim(354\degr, 35\degr)$ and $(l, b)\sim(355\degr, -30\degr)$. The Galactic center spur is visible as the arch in $M_0$ running from about $(0,10)\degr$ to $(350, 25)\degr$.

However, there is more depolarization in the band below $\sim$30$\degr$ around the Galactic plane. This was first identified by \citet{Wolleben2006} in the northern sky in the same wavelength range, partially depolarizing these ridges. If the ridges were formed by outflows of star formation from the Galactic center, the visible parts (at \(|b| \gtrsim 30\degr\)) would be many kiloparsecs above the Galactic plane. As the scale height of thermal free electrons is estimated to be between 1000 pc \citep{Reynolds1991} and 1830 pc \citep{Gaensler2008}, this suggests that at high distances above the plane, the ridges would exhibit negligible electron density and negligible Faraday rotation. In fact, the $M_1$ map shows that the Faraday depth structure of these ridges does not stand out from the background, and $m_2$ shows unresolved spectra in these directions, indicating that the ridges do not give an additional Faraday depth component besides the background. In the data of \citet{Carretti2019}, the ridges are also not visible in the RM map between 2.3 GHz and 22 GHz and 30 GHz, which is confirmed by our conclusion that RM is negligible in these features.

\subsection{Comparison to point source rotation measures}\label{sec:compRM}
 From surveys of polarized extragalactic background sources, \citet{Oppermann2015} derived the Faraday sky map of the Galaxy by removing the intrinsic Faraday rotation component of the background sources, which has recently been improved by \citet{Hutschenreuter2022},\footnote{\url{https://wwwmpa.mpa-garching.mpg.de/~ensslin/research/data/faraday2020.html}} hereafter referred to as $RM_{F.sky}$.
 
\begin{figure}
\center
  \includegraphics[width=8cm]{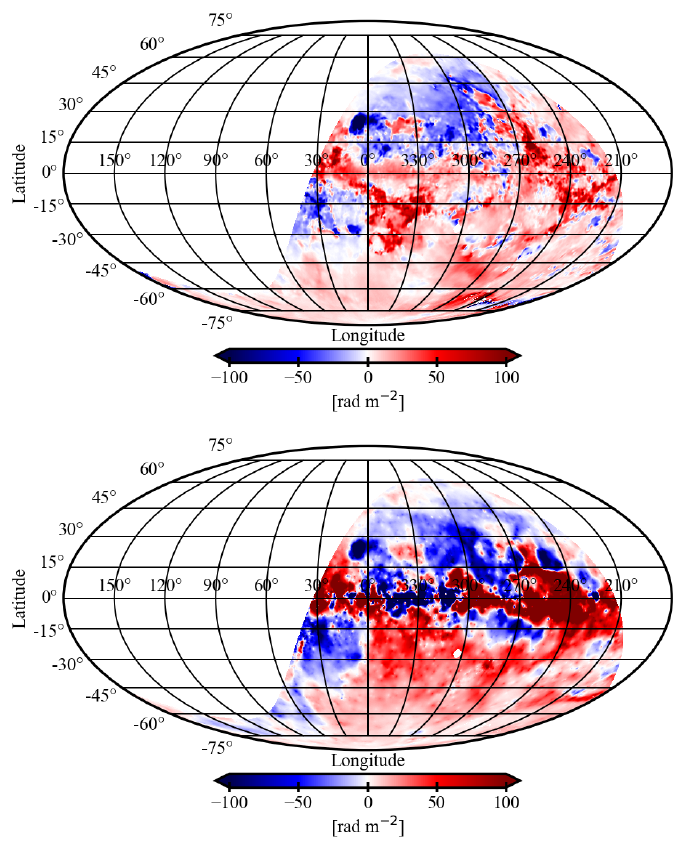}
  \caption{Comparison between $RM_{F.sky}$ and STAPS $M_{1}$ RMs. The top panel: The STAPS $M_{1}$ map smoothed onto the Faraday sky map of \citet{Hutschenreuter2022} at the bottom panel. All the other figure parameters are the same as in Fig.~\ref{fig:staps_moments}).}
  \label{fig:stapsvshutsch}
\end{figure}

 We smoothed STAPS $M_1$ and $RM_{F.sky}$ to a common resolution of $\sim$134~arcmin, chosen as the average difference between data points in the S-PASS/ATCA survey \citep{Schnitzeler2019}, which dominates the southern sky except in the Galactic plane and Magellanic Clouds where the source density is much higher. Fig.~\ref{fig:stapsvshutsch} shows these two maps. We chose to apply this smoothing to the Faraday moments and not to Stokes $Q$ and $U$ to make it as similar as possible to the process applied to obtain the Faraday sky, whereby RM values of point sources were interpolated over the sky.

\begin{figure}
\center
  \includegraphics[width=8cm,height=8cm]{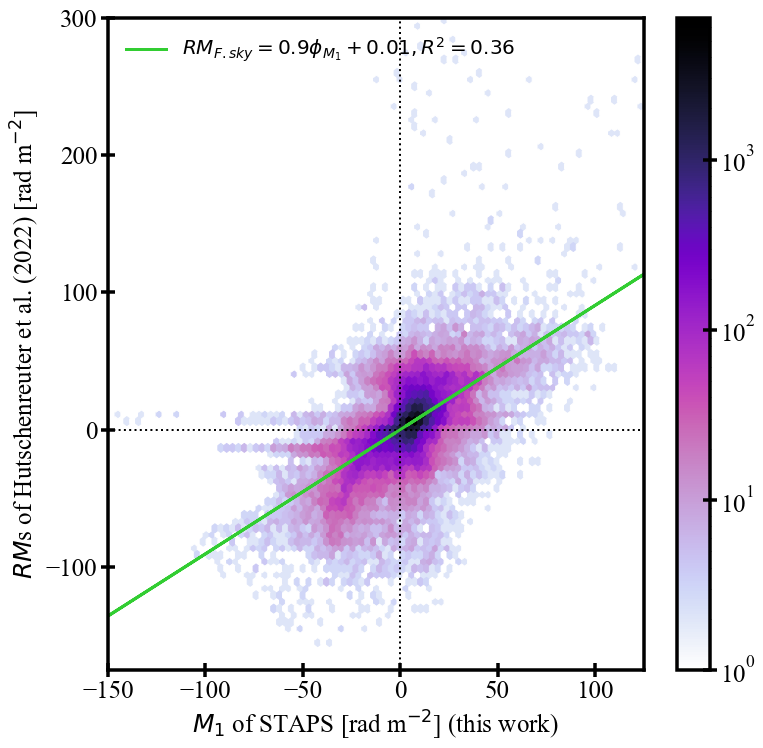}
  \caption{Comparison between $M_{1}$ of STAPS ($\phi_{M_1}$) and the Faraday sky map of \citet{Hutschenreuter2022}. The green line corresponds to the linear fit to the entire data. For display purposes, 10$\times$10 pixels are binned into one larger pixel of 30$\arcmin$. The color bar corresponds to the number of data points in each hexbin. $R$ is the coefficient of correlation.}
 \label{fig:hutsch_scatter}
\end{figure}

We present a comprehensive comparison of the two datasets (STAPS and the Faraday sky of \citealt{Hutschenreuter2022}) using a scatter plot, as is shown in Fig.~\ref{fig:hutsch_scatter}. In both datasets, we masked the following: i) the PI data points with low S/N ($<$6); ii) data points in the Galactic plane with extreme  $RM_{F.sky}$ values and the region of G353$-$34 covering a diameter of $\sim10\degr$; iii) the Sh 2$-$27 H\,{\sc ii} region as there is a very good correlation between $RM_{F.sky}$ and STAPS $M_{1}$ (where the Pearson correlation coefficient, $R$, is 0.74), as is shown in Fig.~\ref{fig:stapsvshutsch}. The remaining data points were then compared with the Faraday sky model.

The Faraday sky map exhibits higher $|RM|$ values than STAPS $|M_{1}|$, which is particularly evident within the Galactic plane. This discrepancy is attributed to the Faraday sky probing the complete LoS through the Galaxy, while the STAPS signal is depolarized at large distances, possibly due to small angular-scale structures in RM. Nevertheless, across most of the high-latitude sky, the features in both maps generally align well. The RMs from extragalactic sources are expected to be twice that of Galactic diffuse emission if the medium behaves as a Burn slab \citep{Burn1966}. However, STAPS $M_{1}$ Faraday depths and the Faraday sky disagree with the expected Burn slab behaviour. Specifically, at intermediate and upper latitudes, the comparison between STAPS $M_1$ and the Faraday sky has a slope of 0.9 and a correlation coefficient of 0.36. This shows that the Burn slab is not a good approximation for the Galaxy, as is pointed out by \citet{Dickey2022}.  However, it is consistent with a model of several emitting and rotating layers, which can produce slopes smaller than 1 \citep{Erceg2022}. Additionally, deviations from the linear relation at high $RM_{F.sky}$ are likely attributed to structures at low latitudes, such as the G353$-$34 region, where the LoS differs between STAPS $M_{1}$ and $RM_{F.sky}$.

\section{Summary and conclusion} \label{sec:conc}
We have presented the Faraday moments observed at 1328$-$1768 MHz from the spectro-polarimetric STAPS survey, also called the GMIMS High-Band South survey, to investigate magnetic structures in the Galaxy. Faraday moments provide a great understanding of radio polarization in the ISM by summarizing the Faraday spectra in the total polarized brightness ($M_0$), the average Faraday depth weighted by the PI ($M_1$), the weighted width of the Faraday spectrum for each pixel ($m_2$), and its skewness ($m_3$). We have used the survey data from STAPS and compared them to relevant previous studies. A comparison of the moment maps with the GHBN survey at the same frequency range in the northern sky shows broad agreement, but also points to significant ground emission in GHBN \citep{Wolleben2021} and edge effects in both surveys.

The conclusions reached in this work are:

\begin{enumerate}
    \item The weighted PI $M_0$ shows a generally low polarization at longitudes of $l \lesssim 300\degr$ in the fourth quadrant, but does show polarization imprints from structures such as the Vela supernova remnant. The S-PASS ridges are visible as high-polarization structures, but only at absolute latitudes higher than the depolarization strip at $-30^{\circ} < b < 30^{\circ}$.
    \item Large parts of the southern sky can be described by single Faraday components, with apparent exceptions being mostly specific structures (supernova remnants, H\,{\sc ii} regions). As a result of the limited frequency range of the survey, Faraday thick structures cannot be detected, or can only be detected at their edges. Therefore, we cannot distinguish between two separate Faraday-thin components in a Faraday spectrum and an edge-detected Faraday-thick structure, and cannot reliably detect Faraday caustics.    
    \item The ratio between the $RM_{F.sky}$ presented by \citet{Hutschenreuter2022} and STAPS $M_{1}$ RMs is 0.9 so that the first moment value is almost equivalent to the total RM value. This indicates a deviation from the expected ideal Burn slab model, but is consistent with a model of multiple emitting elements in a Faraday rotating medium, as is shown by \citet{Erceg2022}.
\end{enumerate}

The Southern Twenty-centimeter All-sky Polarization Survey offers a remarkable opportunity to delve into the intricate characteristics of the Galactic magneto-ionic medium and diffuse sources at frequencies between 1328 and 1768 MHz, with a resolution of 18$\arcmin$. Further research should include exploring individual diffuse sources that could be revealed by STAPS, but also in combination with and comparison to the other GMIMS surveys.

\begin{acknowledgements}
This work is part of the joint NWO-CAS research programme in the field of radio astronomy with project number 629.001.022, which is (partly) financed by the Dutch Research Council (NWO). We thank Karel D. Temmink for the help with coding/parallelization in the astronomy department's coma computing cluster. MH acknowledges funding from the European Research Council (ERC) under the European Union's Horizon 2020 research and innovation programme (grant agreement No 772663). JMS acknowledges the support of the Natural Sciences and Engineering Research Council of Canada (NSERC), 2019-04848. XHS, JLH and XYG are supported by the International Partnership Program of Chinese Academy of Sciences, Grant No. 114A11KYSB20170044. The Dunlap Institute is funded through an endowment established by the David Dunlap family and the University of Toronto. AO is partly supported by the Dunlap Institute. BMG acknowledges the support of the Natural Sciences and Engineering Research Council of Canada (NSERC) through grant RGPIN-2022-03163, and of the Canada Research Chairs program. AB acknowledges financial support from the INAF initiative "IAF Astronomy Fellowships in Italy" (grant name MEGASKAT). MT is supported by the Banting Fellowship (Natural Sciences and Engineering Research Council Canada) hosted at Stanford University and the Kavli Institute for Particle Astrophysics and Cosmology (KIPAC) Fellowship. The authors acknowledge Interstellar Institute's program "With Two Eyes" and the Paris-Saclay University's Institute Pascal for hosting discussions that nourished the development of the ideas behind this work. This work made use of Astropy\footnote{\url{http://www.astropy.org}} a community-developed core Python package for Astronomy \citep{Astrocol2018} and Python modules of Matplotlib\footnote{\url{http://www.matplotlib.org/}} \citep{Hunter2007}, NumPy\footnote{\url{https://numpy.org/}} \citep{Oliphant2006,vanderWalt2011}, and SciPy\footnote{\url{https://www.scipy.org/index.html}} \citep{Virtanen2020}. This research made use of the software package Montage\footnote{\url{http://montage.ipac.caltech.edu/}}. Parts of this work’s results use the color maps in the CMasher package \citep{vanderVelden2020}. The Parkes 64m radio-telescope (\textit{Murriyang}) is part of the Australia Telescope National Facility \footnote{\url{https://ror.org/05qajvd42}} which is funded by the Australian Government for operation as a National Facility managed by CSIRO. We acknowledge the Wiradjuri people as the Traditional Owners of the Observatory site.
\end{acknowledgements}

\end{document}